\documentclass[a4paper]{jpconf}
\usepackage{amssymb,amsmath,amsthm,amsfonts}
\usepackage{graphicx}
\usepackage{hyperref}
\usepackage{tikz}
\usetikzlibrary{decorations.markings,decorations.pathmorphing,arrows}
\usepackage[doublespacing]{setspace}
\usepackage[numbers]{natbib}

\newcommand{\RR}{\mathbb{R}}
\newcommand{\ZZ}{\mathbb{Z}}
\newcommand{\HH}{\mathcal{H}} 
\newcommand{\Al}{\mathcal{A}}
\newcommand{\Inn}[1]{\langle #1 \rangle}

\newcommand\be{\begin{equation}}
\newcommand\ee{\end{equation}}

\begin{document} 

\title{Time-energy uncertainty does not create particles}
\author{B W Roberts$^1$ and J Butterfield$^2$}
\address{$^1$ Philosophy, Logic \& Scientific Method, Centre for Philosophy of Natural and Social Sciences, London School of Economics \& Political Science, London, WC2A 2AE, UK}
\address{$^2$ Trinity College, University of Cambridge, Cambridge, CB2 1TQ, UK}
\ead{\href{mailto:b.w.roberts@lse.ac.uk}{b.w.roberts@lse.ac.uk}, \href{mailto:jb56@cam.ac.uk}{jb56@cam.ac.uk}}
 
\begin{abstract}
  In this contribution in honour of Paul Busch, we criticise the claims of many expositions that the time-energy uncertainty principle allows both a violation of energy conservation, and particle creation, provided that this happens for a sufficiently short time. But we agree that there are grains of truth in these claims: which we make precise and justify using perturbation theory.\\

{\noindent\today. Forthcoming in \emph{Journal of Physics: Conference Proceedings}, Mathematical Foundations of Quantum Mechanics in Memoriam Paul Busch} 
\end{abstract}

\section{Introduction}

In expositions of quantum theory, it is often said that energy conservation can be violated, and that particles can `pop in to existence' out of nowhere, thanks to a time-energy uncertainty principle. Time-energy uncertainty was, of course, one of Paul Busch’s areas of expertise. His insightful analyses of it began already in his two 1990 works \citep{busch1990a,busch1990b} and the topic remained an abiding interest of his, shown for example in his 2008 review article about time in quantum physics \citep{busch2008a}. It seems to us very likely that Busch, with his clarity and precision of thought, would have had misgivings about this ``folklore''. Accordingly, we propose to commemorate his life and work by criticising it---in a way that we hope is worthy of his memory. But we will also argue that the folklore contains grains of truth: which we will make precise and justify, using perturbation theory.

We begin with some illustrative quotations. Thus Jones proposes that,
\begin{quote}
    ``a consequence of the Heisenberg Uncertainty Principle is that we can take seriously the possibility of the existence of energy non-conserving processes—provided the amount by which energy is not conserved, $E_{violation}$, exists for a time less than $t = \hbar/2 E_{violation}$'' \citep[p.226]{jones2002a} 
\end{quote}
This folklore is not just a myth of pedagogy or popular exposition. Many excellent textbooks\footnote{As Fermat might say: we have discovered a marvellous number of comments of this kind, which this footnote is too small to contain. Here are a couple from books about quantum field theory; though one of our main points will be that the issues are not specific to quantum field theory, but arise already in quantum mechanics. The first book is about quantum field theory's philosophical interpretation, the second about its mathematics.   \citet[p.148]{teller-qft}  remarks that violation of local energy conservation is ``customarily excused'' by the time-energy uncertainty principle; while \citet[p.133]{folland2008quantum} writes that, ``the uncertainty principle allows the particle and/or quantum to be temporarily `off mass-shell' between the times of emission and absorption or vice versa''.} make similar comments; such as the classic quantum field theory textbook of Peskin and Schroeder:
\begin{quote}
``Even when there is not enough energy for pair creation, multiparticle states appear, for example, as intermediate states in second-order perturbation theory. We can think of such states as existing only for a very short time, according to the uncertainty principle $\Delta E \cdot \Delta t = \hbar$. As we go to higher orders in perturbation theory, arbitrarily many such `virtual' particles can be created.'' \cite[p.13]{peskinschroeder1995qft}
\end{quote}
Similarly, in their textbook on general relativity, \citet{HobsonEtAl2006gr} say that time-energy uncertainty is responsible not just for virtual particles, but for concrete physical predictions like Hawking radiation\footnote{There is a sense in which Hawking radiation \citep{hawking1975a}, when viewed as a comparison between a quantum field theory constructed at past null infinity and one constructed at future null infinity, is indeed associated with particle creation. However, this is a matter of inequivalent vacua, and so spontaneous symmetry breaking, not of time-energy uncertainty.}:
\begin{quote}
``Hawking's original calculation uses the techniques of quantum field theory, but we can derive the main results very simply from elementary arguments. ... Pair creation violates the conservation of energy and so is classically forbidden. In quantum mechanics, however, one form of Heisenberg's uncertainty principle is $\Delta t\Delta E = \hbar$, where $\Delta E$ is the minimum uncertainty in the energy of a particle that resides in a quantum mechanical state for a time $\Delta t$. Thus, provided the pair annihilates in a time less than $\Delta t = \hbar/\Delta E$, where $\Delta E$ is the amount of energy violation, no physical law has been broken.'' \citep[\S 11.11]{HobsonEtAl2006gr}
\end{quote}

Agreed: not everyone endorses this use of time-energy uncertainty to justify particle creation. In particular, \citeauthor{griffiths1995qm} is unimpressed:
\begin{quote}
  ``It is often said that the uncertainty principle means that energy is not strictly conserved in quantum mechanics-that you're allowed to `borrow' energy $\Delta E$, as long as you `pay it back' in a time $\Delta t \sim \hbar/2\Delta E$; the greater the violation, the briefer the period over which it can occur. There are many legitimate readings of the energy-time uncertainty principle, but this is not one of them. Nowhere does quantum mechanics license violation of energy conservation''. \cite[p.115]{griffiths1995qm}
\end{quote}
We agree with Griffiths.\footnote{Another dissenter is \citet{bunge1970vp}, who gives a short but scathing criticism of understanding virtual particles in terms of time-energy uncertainty.} Of course,  mass-energy is exactly conserved in an isolated physical system, in quantum physics no less than classical physics. In quantum theory, the unitary propagator $U_t$ (a strongly continuous unitary representation of the reals under addition) can be written $U_t = e^{-itH}$, where the self-adjoint generator $H$ is the energy. And, for any initial state $\rho$ that evolves unitarily according to $\rho_t = U_t\rho U_t^*$, the energy expectation value does not change over time, since $H$ and $U_t$ commute: $\Tr(\rho_tH) = \Tr(\rho U_t^*HU_t) = \Tr(\rho H)$ for all $t\in\RR$. 

Nevertheless, what the folklore says about time, energy, and energy conservation contains some grains of truth. We will focus on three ideas:
\begin{enumerate}
  \item (\emph{non-conservation}) There is some sense in which `energy' associated with a perturbed system is not conserved;
  \item (\emph{particle creation}) There is some sense in which that non-conservation allows the non-conservation of particle-number; and  
  \item (\emph{shorter times}) There is some sense in which more particle creation occurs during shorter times. 
  \end{enumerate}
So our aim is to do the exercise of making these statements precise, and verifying them. Our lesson will be that, in each case, it is not a time-energy uncertainty relation that provides the wiggle-room to create particles: hence our slogan, ``time-energy uncertainty does not create particles.'' The particles are rather best viewed as artefacts of the shifted perspective one adopts when approximating a physical system using perturbation theory. We will discuss the statements (i), (ii) and (iii) in Sections \ref{(i)},  \ref{(ii)}  and  \ref{(iii)},   respectively. But we precede this with an overview, Section \ref{ptbnvirtual}, about perturbation theory, and the way it justifies the idea of a virtual state.

\section{The perturbation view of virtual states}\label{ptbnvirtual}

For our purposes, a \textit{quantum system} is a triple $(\HH,\Al,t\mapsto U_t)$, where $\HH$ is a separable complex Hilbert space, $\Al$ is a von Neumann algebra of linear operators on $\HH$, and $t\mapsto U_t$ is a strongly continuous one-parameter unitary representation in $\Al$ of the group $\RR$ under addition. The Hilbert space represents a collection of states; the algebra contains a collection of operators that represent observables; and the unitary representation provides the dynamics. By Stone's theorem, there exists a unique self-adjoint operator $H\in\Al$ (the `energy' operator or `Hamiltonian') such that the dynamics can be written in the form, $U_{t}  =  e^{-itH}$, for all $t\in\RR$.

Perturbation theory uses a quantum system that is in some way tractable to approximate a quantum system that is not. For example: the energy $H$ of a Helium atom, which contains two mutually repulsive electrons, can be approximated by a system whose energy $H_0$ ignores the mutual repulsion of the electrons. In such examples, one begins with a quantum system $(\HH,\Al,t\mapsto U_t)$ and a set of operators $\{A_\lambda\}\subset\Al$ parametrised by a real number $\lambda$, one value of which gives the correct, or physically real,  operator of interest:
\begin{equation}
A_\lambda = A_0 + V_\lambda.
\end{equation}
This set is constructed in such a way that as $\lambda\rightarrow 0$, we get $A_\lambda\rightarrow A_0$ in the operator norm.  $A_0$ is called the `unperturbed' operator, and the set $A_\lambda$ is called a `perturbation'. The hope is both that a physical system can be correctly described by $A_\lambda$ for some value of $\lambda$, and also that its properties can be accurately approximated using known facts about the more tractable operator $A_0$.

Perturbation theory allows one to approximate various aspects of the operator $A_\lambda$ when it can be represented in a power series expansion around $\lambda=0$. That is, one seeks an exact expression of $A_\lambda$ of the form,
\begin{equation}
  A_\lambda = A_0 + \lambda \left(\frac{d}{d\lambda}V_\lambda\right)\Big|_{\lambda=0} + \frac{\lambda^2}{2!}\left(\frac{d^2}{d\lambda^2}V_\lambda\right)\Big|_{\lambda=0} + \cdots
\end{equation}
The $n$th-order approximation of $A_\lambda$ is by definition the sum of the first $n$ terms in this series. As $n\rightarrow\infty$, the series approaches $A_\lambda$ in the operator norm. A wide class of problems can be solved by adopting the simple approximation where only the first two terms are calculated. Defining $V:=\tfrac{d}{d\lambda}V_\lambda\Big|_{\lambda=0}$ this gives what is called a `linear' perturbation:
\begin{equation}
  A_\lambda \approx A_0 + \lambda V.
\end{equation}
The eigenvalue problem for an operator expressed as a linear perturbation can typically be given an approximate analysis thanks to classic results in perturbation theory.\footnote{For example, if it is the case that for all $\lambda$, $A_\lambda = A_0 + \lambda V$ has a discrete spectrum and commutes with its adjoint ($A_\lambda A_\lambda^* = A_\lambda^* A_\lambda$), then by the Mitzkin-Taussky theorem, the eigenvalues of $a^i_\lambda$ of $A_\lambda$ are linear in $\lambda$ (in that $a^i_\lambda = a^i_0 + \lambda k^i$), and its eigenvectors are analytic functions of $\lambda$ \cite[\S 2.5]{kato1995pert}. It is then possible to expand these eigenvectors around $\lambda=0$ to give an approximation in terms of (usually already-known) eigenvectors of $A_0$.}

A typical way for virtual states to arise in perturbation theory is in its application to the dynamics. Let $(\HH,\Al,t\mapsto U_t)$ be a quantum system, and consider a second one-parameter unitary representation $t\mapsto U^0_t$. We write $H$ and $H_0$ for their respective Hamiltonian generators, and refer to the former as the `perturbed' (or `interacting') Hamiltonian, while the the latter is the `unperturbed' Hamiltonian. We define $V:= H-H_0$ and refer to it as the `potential'. Writing $U_t = U^0_t (U^0_{-t}U_t)$, we expand the term in parentheses as a power series around $t=0$,
\begin{equation}\label{seriesforU_t}
    U_t = U^0_t\left( I + t\frac{d}{dt}(U^0_{-t} U_t)\Big|_{t=0} + \frac{t^2}{2!}\frac{d^2}{dt^2}(U^0_{-t} U_t)\Big|_{t=0} + \cdots \right).
\end{equation}
To the extent that $t$ is close to zero and $V=H-H_0$ is small (in the operator norm), the operator $U_t$ is approximated by summing the first $N$ terms in this series and ignoring the rest \cite[\S II.3]{kato1995pert}. Note however that this partial sum of the first $N$ terms is in general not unitary. 

Writing $U^{(n)}_t$ to denote the $n$th term in the series, the first few orders of approximation can be calculated by applying the Leibniz rule to the derivatives:
\begin{align}\label{termsinseriesforU_t}
\begin{split}
    U^{(0)}_t & = U^0_t\\
    U^{(1)}_t & = -itU^0_tV\\
    U^{(2)}_t & = \tfrac{t^2}{2!}U^0_t\left([V,H_0] - V^2\right)
\end{split}
\end{align}

The virtual states picture arises from imagining that each contribution to the series is the amplitude of a ``scattering event'' in its own right. For example, suppose $\psi_a$ and $\psi_b$ are orthogonal eigenvectors of the unperturbed Hamiltonian $H_0$, and that we wish to approximate the amplitude $\Inn{\psi_b,U_t\psi_a}$ associated with a transition during time $t$ from $\psi_a$ to $\psi_b$. In a first-order approximation, we replace $U_t$ with the sum of the terms $U^{(0)}_t + U^{(1)}_t$. In a more accurate second-order approximation, we replace it with $U^{(0)}_t + U^{(1)}_t + U^{(2)}_t$, and so on. With a little calculation\footnote{Zeroth order: $\Inn{\psi_b,U^{(0)}_t\psi_a} = e^{iat}\Inn{\psi_b,\psi_a} = 0$. First-order: $\Inn{\psi_b,U^{(1)}_t\psi_a} = -it  \Inn{\psi_b,U^0_tV\psi_a} = - it e^{ibt}\Inn{\psi_b,V\psi_a}$. Second-order: $\Inn{\psi_b,U^{(2)}\psi_b} = \Inn{\psi_b,U^0_t[V,H_0]\psi_a} - \Inn{\psi_b,U_t^0V^2\psi_a} = e^{ibt}(a-b)\Inn{\psi_b,V\psi_a} + e^{ibt}\Inn{\psi_b,V^2\psi_a}$.}, the terms in these approximations are found to be,
\begin{align}
  \begin{split}
    \Inn{\psi_b,U^{(0)}_t\psi_a} & = 0\\
    \Inn{\psi_b,U^{(1)}_t\psi_a} & = - ite^{ibt}\Inn{\psi_b,V\psi_a}\\
    \Inn{\psi_b,U^{(2)}_t\psi_b} & = e^{ibt}(a-b)\Inn{\psi_b,V\psi_a} - e^{ibt}\Inn{\psi_b,V^2\psi_a}
\end{split}
\end{align}
where $a$ and $b$ are the $H_0$ eigenvalues associated with $\psi_a$ and $\psi_b$, respectively. The zeroth-order contribution $\Inn{\psi_b,U^{(0)}\psi_a}$, viewed as a transition amplitude in its own right, describes an initial state $\psi_a$ that never transitions to $\psi_b$. The first-order contribution might lead us to say that the presence of the potential $V$ `deflects' the initial state $\psi_a$ to $\psi_b$ with some probability. The second-order contribution has two terms, the first of which will just get collected together with the first-order one in the series, and the second of which is a `deflection' by the potential $V^2$. Of course, we have not given any reason to view this language as anything more than short-hand; strictly speaking, each is simply a contribution to a series that approximates $\Inn{\psi_b,U_t\psi_a}$.

Virtual states arise as `intermediate states' in this kind of analysis. In the example just given of approximating the transition amplitude $\Inn{\psi_b,U_t\psi_a}$, they begin to appear at the level of the second-order contribution $U^{(2)}_t$. Instead of thinking of the second-order term $e^{ibt}\Inn{\psi_b,V^2\psi_a}$ as a deflection in the potential $V^2$, let us rewrite it in the form,
\begin{equation}
e^{ibt}\Inn{\psi_b,V^2\psi_a} = \Inn{\psi_b,Ve^{ib(t-t')}\psi'}\Inn{\psi',Ve^{ibt'}\psi_a}
\end{equation}
where we define $\psi':=\tfrac{1}{|V\psi_a|}V\psi_a$, and choose any $t'$ such that $0<t'<t$. Now, instead of viewing the transition as going from $\psi_a$ to $\psi_b$ in the potential $V^2$ during a time $t$, we can view it as consisting of an intermediate transition from $\psi_a$ to $\psi'$ in $V$ during time $t'$, followed by a transition from $\psi'$ to $\psi_b$ in $V$ during the later time-interval $t-t'$. The intermediate state $\psi'$ is an example of a \textit{virtual state}. As expected, the third-order transitions have a $V^3$ term that gives rise to a pair of virtual states, and so on up the series.

One can draw a Feynman diagram for each contribution $\Inn{\psi_b,U^{(n)}_t\psi_a}$ in this series, illustrated for $U^{(0)}$, $U^{(1)}$, and $U^{(2)}$ in Figure \ref{fig:feyn}. Virtual states appear, beginning in the second-order diagrams, as states without open endpoints, such as $\psi'$ in the right-most diagram of the Figure. This encodes the fact that they are not associated with the measured in-state or out-state in the scattering experiment associated with this amplitude. However, again: we have given no reason to view each diagram as anything other than shorthand for a term $\Inn{\psi_b,U^{(n)}\psi_a}$ in a series approximation of the amplitude $\Inn{\psi_a,U_t\psi_b} \approx \Inn{\psi_a,U^{(0)}_t\psi_b} + \Inn{\psi_a,U^{(1)}_t\psi_b} + \Inn{\psi_a,U^{(2)}_t\psi_b} + \cdots$.
\begin{figure}[tbh]\begin{center}
\begin{tikzpicture}
  \begin{scope}[decoration={markings,mark=at position 0.5 with {\arrow{latex}}}]
    \draw[postaction={decorate}] (0,-1.5) -- (0,2.5);
    \node at (0.4,0.5) {$\psi_a$};
  \end{scope}
\end{tikzpicture}
\hspace{3cm}
\begin{tikzpicture}
\begin{scope}[decoration={markings,mark=at position 0.5 with {\arrow{latex}}}]
  \draw[postaction={decorate}] (1,-1.5) -- (0,0);
  \draw[postaction={decorate}] (0,0) -- (0,2.5);  
  \draw[decorate,decoration={snake, segment length=6pt, amplitude=1pt}]  (-1,-1.5) -- (0,0);
  \node at (-0.75,-0.6) {$V$};
  \node at (0.9,-0.6) {$\psi_a$};
  \node at (0.4,1.1) {$\psi_b$};
\end{scope}
\end{tikzpicture}
\hspace{2cm}
\begin{tikzpicture}
\begin{scope}[decoration={markings,mark=at position 0.5 with {\arrow{latex}}}]
  \draw[postaction={decorate}] (1,-1.5) -- (0,0);
  \draw[postaction={decorate}] (0,0) -- (0,1);
  \draw[postaction={decorate}] (0,1) -- (1,2.5);
  \draw[decorate,decoration={snake, segment length=6pt, amplitude=1pt}]  (-1,-1.5) -- (0,0);
  \draw[decorate,decoration={snake, segment length=6pt, amplitude=1pt}]  (0,1) -- (-1,2.5);  
  \node at (-0.75,-0.6) {$V$};
  \node at (-0.75,1.5) {$V$};
  \node at (0.75,1.5) {$\psi_b$};    
  \node at (0.9,-0.6) {$\psi_a$};
  \node at (0.35,0.5) {$\psi'$};
\end{scope}
\end{tikzpicture} 
\caption{}\label{fig:feyn} Feynman diagrams for the first three terms in a perturbation series for $\Inn{\psi_b,U_t\psi_a}$, with $U_t = e^{-it(H_0 + V)}$ and $\psi_a,\psi_b$ eigenvectors of $H_0$. The second-order state $\psi'$ is `virtual'.
  \end{center}\end{figure}

\section{The appearance of energy non-conservation}\label{(i)} 
To sum up: the perturbation view is one of shifting perspectives. We describe a quantum system $(\HH,\Al,t\mapsto U_t)$ from the perspective of the $n$th-order approximation in a perturbation series, recognising that on this perspective, the system will sometimes appear to deviate from its true `perturbed' dynamics, as well as from the idealised `unperturbed' dynamics. 

As it happens, one such deviation is that energy conservation can appear to fail. Thus we have our first claim:
\begin{enumerate}
\item (\emph{non-conservation})  There is some sense in which `energy' associated with a perturbed system is not conserved.
\end{enumerate}
In the context of the virtual states described in the previous section, the `energy' that fails to be conserved is associated with the idealised Hamiltonian $H_0$ of the unperturbed dynamics. In the zeroth-order approximation $U^{(0)}$, it is conserved. However, in the first-order and second-order approximations, it is not, in that there are eigenstates of $H_0$ that are not stationary. This is due to the fact that for interacting systems with $[H_0,V]\neq0$, the unperturbed Hamiltonian is not stationary, $U_tH_0U_t^*\neq0$, or equivalently, $[H,H_0]\neq0$. So this is one thing that could be meant  by saying `energy' is not conserved in the presence of virtual states.

However, in a quantum system $(\HH,\Al,t\mapsto U_t)$, it is the generator $H$ of the true i.e. perturbed dynamics $U_t$ that represents the `true energy' of the system, not the idealised Hamiltonian $H_0$. So, a more physically interesting question is whether $H$ commutes with the approximate dynamics given by a partial sum, up to say the $N$th term, of the perturbation series, eq. \ref{seriesforU_t} and \ref{termsinseriesforU_t}.

Of course, it may happen that for some finite $N$, we find that $\sum^N_n U^{(n)} = U_t$, so that $\sum^N_nU^{(n)}$ is an exact description rather than an approximate one, and the energy $H$ is indeed conserved. But in general, the approximate dynamics $\sum^N_nU^{(n)}$ is not equal to $U_t$, indeed is not even unitary; and $H$ will not be conserved under it.
 
On the other hand, there is a sense in which energy $H$ for the perturbed dynamics $U_t$ comes closer to being conserved by the approximate dynamics, the higher the order  of approximation. For as we add more terms, the resulting approximate dynamics $\sum_{n}U^{(n)}_t$ better approximates the true dynamics $U_t$; that is, $\sum_{n=0}^NU^{(n)}_t\rightarrow U_t$ in the operator norm as $N$ becomes arbitrarily large. But this implies:\footnote{In general, $\|A_n - B\|\rightarrow 0$ implies the commutator $\|[A_n,B]\|\rightarrow 0$, since
  $\|[A_n,B]\| = \|(A_n - B)B + B(B-A_n)\| \leq \|A_n - B\|\|B\| + \|B\|\|B - A_n\| = 2\|B\|\|A_n - B\|$.}
\begin{equation}
  \left[\textstyle\sum_{n=0}^NU^{(n)}_t,H\right]\rightarrow0 \text{ as } N\rightarrow+\infty,
\end{equation}
giving perfect conservation of energy in the limit. This is ironic, in that \citeauthor{peskinschroeder1995qft} write: ``As we go to higher orders in perturbation theory, arbitrarily many such `virtual' particles can be created'', suggesting that energy conservation gets worse with higher order terms \cite[p.13]{peskinschroeder1995qft}. Agreed: there is `more room' for virtual states in higher-order terms. But this does not mean that energy conservation gets worse. In general, it gets better as the `true' perturbed energy becomes more distant (in the operator norm) from the idealised, unperturbed description of the system. 

\section{Particle creation}\label{(ii)}

We now turn to the statement of (ii), the particle-number claim. For this we need to add some notion of particle number to our description. This will consist in a representation of annihilation ($a_i$) and creation ($a_i^*$) operators on $\HH$ for $i\in\ZZ^+$, which satisfy $[a_i,a_j^*] = \delta_{ij}$ and $[a_i,a_j] = [a_i^*,a_j^*] = 0$. Let $N = \sum_ia_i^*a_i$ be the `particle number' operator. Taking $U^0_t$ as the unperturbed dynamics associated with no interactions, we assume for simplicity that $[N,U_t^0]=0$, and hence that the unperturbed system is one in which this (unperturbed) particle-number is conserved. However, if $[N,U_t]\neq0$, then particle number will in general \emph{not} be conserved by the `true'  dynamics; and in general, it will not be conserved under the dynamics given by any of the partial-sum approximations $\sum_nU^{(n)}_t$. (Agreed, $N$ {\em might} be conserved by $U_t$ and-or by an approximate dynamics. But this will not hold true in general.)

That is, we have:
\begin{enumerate}
\setcounter{enumi}{1}
\item (\emph{particle-creation}) The dynamics generated by the partial-sum approximation $\sum_n U^{(n)}_t$ does not in general conserve the unperturbed particle number $N$.
\end{enumerate}

\section{Shorter times}\label{(iii)} 

We now turn to our final claim:
\begin{enumerate}
\setcounter{enumi}{2}
\item (\emph{shorter times}) There is some sense in which more particle creation occurs during shorter times. 
\end{enumerate}
Here  we enter the realm of the time-energy uncertainty principle---or, rather, principles---that are invoked by the cavalier textbook tradition with which we began. These principles are surveyed by \citet{busch1990a,busch1990b,busch2008a}. For us, there are two main points to make, corresponding to two broad understandings of time-energy uncertainty. The first point is not connected to perturbation theory: it is really a warning against an untenable understanding of time-energy uncertainty. The second point will be more positive, in that it will vindicate the shorter-time claim (iii). 

The first point concerns what \citet{busch2008a} suggests one should call ‘external time’ (or in \cite{busch1990a}: `pragmatic time'): namely, time as measured by clocks that are not coupled to the objects studied in the experiment. So in this role, time specifies a parameter or parameters of the experiment: e.g. an instant or duration of preparation or of measurement, or the time-interval between preparation and measurement. In this role, there  seems to be no scope for uncertainty about time. And indeed, our first point here is a warning---following \citet{busch1990a}. For as Busch discusses, there is a tradition (deriving from the founding fathers of quantum theory) of an uncertainty principle between:
\begin{enumerate}
  \item[(1)] the duration of an energy measurement; and
  \item[(2)] either the range of an uncontrollable change of the measured system’s energy, or the resolution of the energy measurement, or the statistical spread of the system’s energy.
\end{enumerate}

To give a little more detail: \citet[\S1-2]{busch1990a} describes how various authorities (including \citet{landau1931erweiterung} and \citet[\S 44]{LandauLifshitz1958qm}) endorse at least one of the following claims:
\begin{enumerate}
  \item[($P$)] An energy measurement of duration $\Delta t$ leads to an uncontrollable and unpredictable change of the (previously sharply defined) energy by an amount of the order  $\Delta E$ such that $\Delta E. \Delta t \geq \hbar$; so that there is no short-time reproducible (first kind) energy measurement.
  \item[($P'$)] An energy measurement of duration $\Delta t$  must carry an inaccuracy $\Delta E$  such that the uncertainty relation $\Delta E. \Delta t \geq \hbar$ is satisfied.
\end{enumerate}
Busch argues, and we agree, that \citet{aharonovbohm1961a} refute this tradition; (see \cite{busch1990a}, especially \S 4, and \cite[\S 3.1]{busch2008a}). They give a simple model of an arbitrarily accurate and arbitrarily rapid energy measurement, where in short: two particles are confined to a line and are both free, except for an impulsive measurement of the momentum and so energy of the first by the second, with the momentum of the second being the pointer-quantity.\footnote{Note that Busch argues that a proper analysis and vindication of Aharonov and Bohm’s refutation uses positive operator-valued measures (POVMs) to describe measurement outcomes, a notion of physical quantity that generalises projection-valued measures (PVMs); this follows the tradition of \citet{ludwig1983a,ludwig1985a}, and is developed at book-length by \citet{BuschEtAl1995-OpQuantPhys}.} This leaves little room for `energy uncertainty' to develop during brief (in external time) energy measurements, either in terms of uncontrollable changes in energy, or in terms of measurement inaccuracy.

Our second, more positive, point concerns what \citet{busch2008a} suggests  one should call  `intrinsic time' (or in his \cite{busch1990a}: `dynamical time'): namely, a dynamical variable of the target system itself whose function is to measure the time, such as the position of a clock's dial relative to its face. Busch suggests that in principle, every non-stationary quantity $A\in\Al$ and density operator state $\rho$ defines a characteristic time interval, $\tau_{\rho}(A)$, in which the expectation value changes `significantly'. For example: in the Schr\"odinger representation on the space of $L^2(\RR)$ functions, if $A\psi(x): = Q\psi(x) = x\psi(x)$ for all $\psi(x)$ in the domain of $Q$ and if $\rho = E_\psi$ is the projection associated with a wave packet $\psi(x)$, then $\tau_{\rho}(A)$ could be defined as the time interval required for the bulk of the wave packet to shift by its width---in some sense of `width'.

Various definitions of such a time interval are available. We will choose what is sometimes called the `characteristic time' associated with the dispersion of an operator-state pair. This obeys what is probably the best-known time-energy uncertainty principle for intrinsic times: the {\em Mandelstam-Tamm uncertainty principle}. 

One arrives at it for a quantum system $(\HH,\Al,t\mapsto U_t)$ by combining three ideas: (a) the Heisenberg equation of motion for an operator $A\in\Al$ with $A(t):=U_tAU^*_t$,
\be
i \hbar \tfrac{d}{dt}A(t) = [H,A(t)];
\label{Heqmot}
\ee
(b) the Heisenberg-Robertson uncertainty principle\footnote{Cf. \cite[Theorem 8.1.2]{BlankExnerHavlicek}}, that for any quantities $A, B\in\Al$, and density operator 
(quantum state) $\rho$,
\be
\Delta _{\rho}A \Delta _{\rho}B \geq  \frac{1}{2} | \Tr(i[A,B]\rho) | \;\; ;
\label{RobUP}
\ee
and (c) the definition of a characteristic time for an operator $A$ that does not commute with the unitary dynamics $t\mapsto U_t$, and for a time-independent state $\rho$ that is not an eigenstate of $A$:
\be
\tau_{\rho}(A) := \frac{\Delta_{\rho} A}{|\tfrac{d}{dt}\Tr(A(t)\rho)|},
\label{tauMT}
\ee
i.e. the time it takes for the expectation value of $A$ to change by its standard deviation. From these definitions it immediately follows that,
\be
 \tau_{\rho}(A) \Delta_{\rho}(H) \geq  \frac{1}{2} \hbar,
 \label{MTUP}
 \ee
whenever $[A,H]\neq 0$ and $\rho$ is not an eigenstate of $H$.
 
This principle can be applied in a perturbation theory analysis of a quantum system $(\HH,\Al,t\mapsto U_t)$ with Hamiltonian generator $H$ and an unperturbed Hamiltonian $H_0$, in two ways, so as to give two construals of the shorter times claims (iii). The first way puts $H_0$ for $A$ in eq. \ref{MTUP}, while the second way puts a number operator $N$ for $A$. 

So first set $A := H_0$ in eq. \ref{MTUP}, to get: $\tau_\rho(H_0)\Delta_{\rho}(H) \geq \tfrac{1}{2}\hbar$.  Suppose that $\Delta_\rho H_0$ is very small, corresponding to a state $\rho$ that is `peaked' in unperturbed energy, so that the characteristic time $\tau_\rho(H_0)$ is comparatively small. The Mandelstam-Tamm uncertainty principle then implies that $\Delta_\rho H$ is comparatively large, and hence that the spread of the system's true (unperturbed) energy is large. In this sense, a short characteristic time and a peaked distribution of the unperturbed energy is associated with a large uncertainty of the system's true energy. Thus we have one general way to construe the shorter-times claim (iii) above. Agreed, this is a construal that uses the unperturbed Hamiltonian $H_0$, not an (unperturbed) number operator $N$.

And finally: a related construal of claim (iii)  can be given, about a particle number operator $N$ associated with the unperturbed dynamics i.e. that satisfies $[N,H_0]=0$; so that under unperturbed dynamics, this particle number is conserved. Such an operator will  typically not be conserved by the true dynamics, in that $[H,N]\neq0$. Thus let us suppose now that $\rho$ is a state for which the true i.e. perturbed energy $H$ is peaked, in that $\Delta_\rho H$ is small, and hence (eq. \ref{MTUP}) that the characteristic time (under the true i.e. perturbed dynamics) $\tau_\rho(N)$ is large. Then, the larger  $|\tfrac{d}{dt}\Tr(N(t)\rho)|$ is, corresponding to a fast rate of change of the expectation value of $N$ in the true dynamics, the larger the spread $\Delta_p N$ must be, in order to guarantee that $\tau_\rho(N):=\Delta_\rho N/|\tfrac{d}{dt}\Tr(N(t)\rho)|$ (cf. eq. \ref{tauMT}) is sufficiently large. So this construal of claim (iii) amounts to: `If the perturbed energy $H$ is peaked, and the expectation value of $N$ changes fast (``short times''), then $N$ has large spread (``non-negligible amplitudes for values far from the expectation value'')'.\\

{\noindent To conclude: we hope to have shown that with a little {\em Buschian wisdom}, one can recover some grains of truth from some cavalier statements in the textbook tradition. May Paul's legacy continue to inspire our community to emulate his craftsmanship and creativity.}

\ack{We thank Klaas Landsman, Jos Uffink and Reinhard Werner for advice and encouragement at an early stage; two anonymous referees for helpful comments and corrections; and the editors, not least for their patience. Bryan W. Roberts was supported by the Leverhulme Trust's Philip Leverhulme Prize and as a Visiting Fellow Commoner at Trinity College, Cambridge.}

\newcommand{\newblock}{} 

\end{document}